\documentclass[conference]{IEEEtran}
\IEEEoverridecommandlockouts
\usepackage{cite}
\usepackage{amsmath,amssymb,amsfonts}
\usepackage{algorithmic}
\usepackage{graphicx}
\usepackage{tikz}
\usepackage{circuitikz}
\usepackage{caption}
\usepackage{subcaption}
\usepackage{textcomp}
\usepackage{xcolor}
\usepackage{booktabs}
\usepackage{pgfplots}
\usepackage{siunitx}
\def\BibTeX{{\rm B\kern-.05em{\sc i\kern-.025em b}\kern-.08em
    T\kern-.1667em\lower.7ex\hbox{E}\kern-.125emX}}
    
\begin{document}

\title{Optimizing Experiments for Accurate Battery Circuit Parameters Estimation: Reduction and Adjustment of Frequency Set Used in Electrochemical Impedance Spectroscopy\\
}

\author{\IEEEauthorblockN{Vladimir Sovljanski, Mario Paolone}
\IEEEauthorblockA{\textit{Distributed Electrical Systems Laboratory} \\
\textit{\'Ecole Polytechnique F\'ed\'erale de Lausanne}\\
Lausanne, Switzerland \\
\{vladimir.sovljanski\}, \{mario.paolone\}@epfl.ch}
\and
\IEEEauthorblockN{Sylvain Tant, Damien Pierre Sainflou}
\IEEEauthorblockA{\textit{Stellantis} \\
Carri\`{e}res-sous-Poissy, France \\
\{sylvain.tant1\}, \{damienpierre.sainflou\}@stellantis.com}
}

\maketitle

\begin{abstract}
In this paper, we study a suitable experimental design of electrochemical impedance spectroscopy (EIS) to reduce the number of frequency points while not significantly affecting the uncertainties of the estimated cell's equivalent circuit model (ECM) parameters. It is based on an E-optimal experimental design that aims to maximize the information about the ECM parameters collected by EIS measurements and, at the same time, minimize the overall uncertainty. In a numerical experiment, we first analyze to which extent reducing the number of measurement points at low frequencies affects the uncertainty of the estimated parameters. Secondly, we show that applying the frequency adjustments can lead to the same or even improved global uncertainty of ECM parameter estimates as with a higher number of measurements. This is numerically verified through a case study using the ECM parameters of a commercial battery cell.
\end{abstract}


\section{Introduction}\label{sec:Intro}
The number of electric vehicles (EVs) has been increasing exponentially. According to estimates, by 2030, the proliferation of EVs will result in the availability of 100–200 GWh/year of electricity storage through EVs \cite{zhu_end--life_2021}. 
After being used in EVs under highly dynamic charging and discharging profiles and reaching a particular capacity level (e.g., 80\% of the beginning of life (BoL) capacity), EV battery cells can be repurposed, in their second life, for less-demanding applications such as stationary energy storage systems (ESSs) \cite{kotak_end_2021}, \cite{gladwin_viability_2013}.

Electrochemical impedance spectroscopy (EIS) is a powerful, non-invasive technique for characterizing electrochemical systems, including batteries. Together with equivalent circuit models (ECMs), it is widely used to characterize the impedance and estimate parameters of battery cells \cite{iurilli_use_2021}. Identifying these parameters is necessary for the characterization of battery cells, ensuring they meet safety and operational standards \cite{european_commission_joint_research_centre_sustainability_2018}.

EIS is carried out to study electrochemical processes occurring at different dynamic rates inside the cell. To capture low-frequency (LF) behaviour, i.e., diffusion processes, the EIS must be performed at frequencies as low as 0.01 Hz, which requires substantial experimental time. According to \cite{iurilli_use_2021}, this is one of the reasons why the impedance in the LF region is often neglected or even not measured. 

Large measurement time at LF can be reduced by using excitation signals with multiple sinusoidal components, which reduces the experimental time but often results in cross-coupling of the cell's response and difficulties in isolating the input impedance for each component \cite{lazanas_electrochemical_2023}. EIS instrumentation manufacturers offer the possibility to perform the EIS using multisine signals \cite{gamry_optieis_nodate}, \cite{biologic_eis_nodate}. Authors of \cite{ciucci_reducing_2011} propose tools for reducing experimental time and errors of 3-4 estimated parameters of solid-state electrochemical cells model from EIS measurements. 

Regarding the accuracy of ECM estimated parameters, in \cite{sovljanski_use_2024}, we analyze how to improve the variance of ECM parameters estimated from EIS measurements through the Cram\'{e}r-Rao Lower Bound (CRLB) and propose an algorithm for optimal frequency adjustments based on E-optimal design that provides the improvement of the overall estimation accuracy.
In this paper, we study how to reduce and adjust the frequency set at which one performs the EIS while characterizing a Li-ion cell to maintain the global accuracy of performing the EIS with a higher number of frequency points.

The paper is organized as follows: in Section \ref{sec:method}, we present the theory on ECM modelling of Li-ion cells, parameters estimation and their accuracy assessment, EIS experimental time and recall the steps for optimal frequencies adjustments. In Section \ref{sec:numerical_study}, we present the results of a case study which analyses the impact of frequency set reduction on estimated parameters accuracy and shows that a suitable adjustment of a reduced set of frequencies can bring the overall ECM estimated parameters accuracy back to the same level as the estimation using a higher number of points per decade (PPD) at certain frequencies. Section \ref{sec:conclusion} concludes the paper.

\section{Method}\label{sec:method}

\subsection{Choice of ECM}

Once the EIS spectrum is obtained in the Nyquist plot, the modeller examines it and chooses a suitable ECM that may correspond to the spectrum \cite{srinivasan_introduction_2021}. Typical ECMs to model Li-ion batteries usually consist of elements modelling high-, mid- and low-frequency phenomena occurring inside the cell. The high-frequency (HF) is typically modelled with a series resistance connected to an inductive element. The inductive element can be a pure inductance if the spectrum at HF is orthogonal to the real axis in the complex plane. However, due to the potential presence of skin effect, one can notice a slight inclination of the spectrum. In this case, to improve the fitting, it is preferable to use a constant phase element (CPE) whose complex impedance is given as 
\begin{equation}
    \bar{Z}_\text{CPE} = \dfrac{1}{Q\cdot (j\omega)^\phi}
\end{equation}
with the exponent $\phi \in \left[-1,0\right)$. The mid-frequency (MF) is usually modelled with one or multiple Zarc elements (parallel connection of a resistor and a CPE), connected in series. The number of Zarc elements can be decided by inspecting the curvatures of the MF part of the spectrum. Finally, the LF is modelled using the Warburg impedance, which is the special case of CPE with $\phi = 0.5$, i.e., it produces the unity slope in the Nyquist plot. Similarly to HF, if the slope deviates from the unity, instead of pure Warburg impedance, one can improve the fitting by using an LF CPE element, with $\phi \in \left[0,1\right)$, however usually close to $0.5$.

\begin{figure*}[!h]
\begin{center}\begin{circuitikz}[scale = 0.8]
    \draw[
        red,
        dashed,
        fill=red!5
    ] (2.35-4.65,1.15) rectangle (8.15-5.85,4.85) node[midway,below=1.5cm,black] {HF region};
    \draw[
        teal,
        dashed,
        fill=green!5
    ] (2.38,1.15) rectangle (8.15,4.85) node[midway,below=1.5cm,black] {MF region};
    \draw[
        blue,
        dashed,
        fill=blue!5
    ] (2.35+5.9,1.15) rectangle (8.15+2.2,4.85) node[midway,below=1.5cm,black] {LF region};
   \draw
  (-2.5,3) to[R, n=Rs] ++(2.5,0) (Rs.s) node[below] {$R_s$}
  (-2.5,3) node[circ]{}
  (0,3) node[circ]{}
  (0,3) to[european resistor=$\bar{Z}_{HF}(\omega)$, n=ZHF] ++(2.5,0) (ZHF.s) node[below] {$Q_{HF},\phi_{HF}$}
  (2.5,3) node[circ]{}
  (2.5,4) to[european resistor=$\bar{Z}_{CPE,1}(\omega)$, n=ZCPE1] ++(2.5,0) (ZCPE1.s) node[below] {$Q_{1},\phi_{1}$}
  (2.5,4) to[short] ++(0,-2)
  to[R, n=R1] ++(2.5,0) (R1.s) node[below] {$R_1$}
  (5,3) node[circ]{}
  (5,3) to[short] ++(0.5,0)
  (5.5,3) node[circ]{}
  (5,4) to[short] ++(0,-2)
  (5.5,4) to[european resistor=$\bar{Z}_{CPE,2}(\omega)$, n=ZCPE2] ++(2.5,0) (ZCPE2.s) node[below] {$Q_{2},\phi_{2}$}
  (5.5,4) to[short] ++(0,-2)
  to[R, n=R2] ++(2.5,0) (R2.s) node[below] {$R_2$}
  (8,4) to[short] ++(0,-2)
  (8,3) to[european resistor=$\bar{Z}_{LF}(\omega)$, n=Zw] ++(2.5,0) (Zw.s) node[below] {$Q_{LF}, \phi_{LF}$}
  (8,3) node[circ]{}
  (10.5,3) node[circ]{};
\end{circuitikz}\end{center}
\caption{ECM modelling the Li-ion cell impedance.}
\label{fig:FullRandlesCircuiTikZ}
\end{figure*}
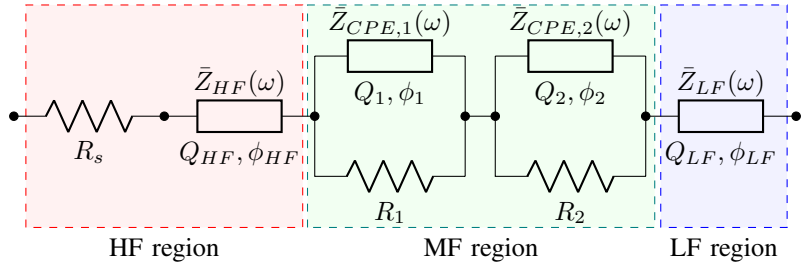

In this work, we model the battery cell with ECM shown in Fig.~\ref{fig:FullRandlesCircuiTikZ}, which represents a generalized Randles equivalent circuit. It includes a series resistance $R_s$, constant phase element (CPE) at high-frequencies (HF) with parameters $Q_{HF}$ and $\phi_{HF}$, two Zarc elements at MF connected in series with parameters $Q_i$ and $\phi_i$, $i\in\left\{1,2\right\}$  and CPE at LF with parameters $Q_{LF}$ and $\phi_{LF}$. Choosing CPEs in all HF, MF, and LF regions allows us to improve the fitting. All ECM parameters are included in the generic vector $\boldsymbol{\theta}$.

\subsection{Parameters Estimation}
As known, EIS is carried out at multiple different and used-defined frequencies, $f_i = \omega_i/(2\pi)$, covering the entire frequency range of interest by exciting the cell with a small sinusoidal signal that also generates a sinusoidal response in a pseudo-linear regime \cite{wang_electrochemical_2021}. After performing the EIS, ECM parameters are estimated from the impedance data points $\tilde{Z}(\omega_i)$.

The common method for impedance data fitting is the Weighted Complex Non-linear Least Squares (WCNLS) \cite{orazem_electrochemical_2017}. It consists of fitting the complex impedance measurements,  $\tilde{Z}(\omega_i)$, to the model function  $Z(\boldsymbol{\theta},\omega_i)$ representing the total ECM impedance. In polar coordinates, the WCNLS simultaneously fits impedance modules and phases. Assuming that impedance's magnitude and phase measurements are uncorrelated, parameters are estimated by solving the following unconstrained minimization problem:
\begin{equation}
     \hat{\boldsymbol{\theta}} = \underset{\boldsymbol{\theta}}{\text{argmin}} \sum_{i=1}^{N} \tfrac{1}{\sigma^2_{\rho}(\rho_i)}(\tilde{\rho}_i - {\rho}_i(\boldsymbol{\theta}))^2 + \tfrac{1}{\sigma^2_{\varphi}}(\tilde{\varphi}_i - {\varphi}_i(\boldsymbol{\theta}))^2
     \label{lab:objective_fun}
\end{equation}
  where the quadratic mismatches between magnitudes, $\tilde{\rho}_i = |\tilde{Z}(\omega_i)|$ and ${\rho}_i(\boldsymbol{\theta}) = |Z(\boldsymbol{\theta},\omega_i)|$ and phases, $\tilde{\varphi}_i = \arg{\tilde{Z}(\omega_i)}$ and ${\varphi}_i(\boldsymbol{\theta}) = |Z(\boldsymbol{\theta},\omega_i)|$ are weighted using the corresponding variances of measured magnitudes and phases. These are extracted from the measurement model, i.e., the error structure provided by the EIS instrumentation manufacturer. Due to the  non-convex objective function, the ECM parameters are first initialized by leveraging the properties of the EIS spectrum and geometric shapes each of the elements produces, as proposed in \cite{sovljanski_use_2024}.

\subsection{On the Uncertainties of the Parameters Estimation}\label{subsec:uncertainties}
To find the least possible variance of an unbiased estimator, defined by the Cram\'{e}r-Rao Lower Bound, we first compute the Fisher Information Matrix (FIM). This is done under the assumption that the mismatches between the measurements and measurement model follow Gaussian distribution with zero mean and a certain variance that can be computed using the information about measuring device precision. We then find the corresponding log-likelihood function in terms of model parameters and its partial derivatives with respect to each parameter. These are then used to compute the elements of the FIM. The $(k,l)-$th element is given by the following compact expression \cite{kay_fundamentals_1993}: 
\begin{multline}\label{eq:CRLB_exp}
[ \boldsymbol{\mathcal{F}}(\boldsymbol{\theta},\boldsymbol{\omega})]_{k,l} =  \dfrac{\partial{\mathcal{Z}}^\top(\boldsymbol{\theta},\boldsymbol{\omega})}{\partial\theta_k} \boldsymbol{Q}^{-1}(\boldsymbol{\theta})\dfrac{\partial{\mathcal{Z}}(\boldsymbol{\theta},\boldsymbol{\omega})}{\partial\theta_l}\\
+\dfrac{1}{2}\text{tr}
\left(
\boldsymbol{Q}^{-1}(\boldsymbol{\theta})
\cdot\dfrac{\partial \boldsymbol{Q}(\boldsymbol{\theta})}{\partial\theta_k}
\cdot \boldsymbol{Q}^{-1}(\boldsymbol{\theta})
\cdot \dfrac{\partial \boldsymbol{Q}(\boldsymbol{\theta})}{\partial\theta_l}
\right)
\end{multline}
where $\mathcal{Z}(\boldsymbol{\theta},\boldsymbol{\omega})$ contains collection of model functions for all $\omega_i$, and $\boldsymbol{Q}$ is the covariance matrix.
After evaluating the FIM elements at the true parameter values\footnote{True parameters are unknown in practice. Therefore, the CRLB is calculated using the estimated parameters' values. Throughout numerical experiments, where we have access to the values of the true parameters, it is confirmed that such calculation does not significantly impact the algorithm's decisions and results, as shown in \cite{sovljanski_use_2024}.}, $\boldsymbol{{\theta}_{\text{true}}}$, the diagonal elements of its inverse, provide CRLB for each parameter.

The CRLB is highly dependent on the frequency set at which one performs the EIS measurements. As known from the information theory, higher number of measurement points lead to more information about model parameters, and therefore results in lower uncertainty in estimates. In addition, measurements at different frequencies contribute differently to different ECM parameters, e.g., increasing the number of impedance measurements at LF higher affects LF than HF ECM parameters.

\subsection{EIS Experimental Time}
EIS is conventionally performed at logarithmically-spaced frequencies, from initial $ f_{start} $ to final $ f_{end} $ with a certain number of $ \text{PPD} $ starting at high frequencies, i.e., $ f_{start}>f_{end} $. The intermediate frequencies are usually selected according to \cite{gamry_reference_nodate}:
\begin{equation}
    f_k = 10^{\log_{10}(f_{start}) - \frac{k}{\text{PPD}}},\;\;\; k = \left\{0,1,\dots,N-1\right\}
\end{equation}
where $ N = \lfloor 1.5+\text{PPD}\cdot(\log_{10}f_{start} - \log_{10}f_{end}) \rfloor $ is the total number of measurement points. The impedance is extracted by exciting the cell with $N_p$ periods of a sinusoidal signal, where $ N_p $ is usually three to five. Therefore, the total experimental time to perform the EIS can be calculated as $ t_{tot} = \sum_{k=0}^{N-1}\frac{N_p}{f_{k}}.$
Measuring at LF requires a great fraction of the total experimental time. For instance, performing the EIS from $f_{start} = 10$ kHz to $f_{end} = 0.01$ Hz with $\text{PPD} = 10$ and $N_p = 5$ results in $t_{tot} = 36.9 $ min, where the two lowest decades, from $0.01$ Hz to $0.1$ Hz and from $0.1$ Hz to $1$ Hz exhaust 90\% and 9\% of the total experimental time, respectively. In this work, we study the possibility of reducing the number of $\text{PPD}$ only at frequencies below $1$ Hz (or $0.1$ Hz) and then performing the frequency adjustments to show that achieving the same overall uncertainty of estimated parameters is possible while reducing the total experimental time.  
 
\subsection{E-optimal Design for Frequencies Adjustments with reduced Number of PPD at LF}

\color{black}{In general, the objective of E-optimal design is minimizing the largest eigenvalue of the FIM. Consequently, it minimizes the smallest eigenvalues of its inverse, so-called dispersion matrix. These eigenvalues represent the square root of lengths of associated uncertainty ellipsoid's axes \cite{fedorov_model-oriented_1997}.} The ellipsoid's volume is proportional to the square root of the product of the eigenvalues of the FIM inverse, $\boldsymbol{\mathcal{F}}^{-1}$.
\color{black}

In \cite{sovljanski_use_2024}, we introduced an algorithm which adjusts the frequencies for the EIS, resulting in improved accuracy in estimated ECM parameters. Initially, the EIS is performed at log-spaced frequencies with a certain number of PPD. 
Measuring at this set of frequencies provides a certain minimum level of uncertainty given by the CRLB. Due to the high experimental time required to perform the EIS at LF, we first reduce the number of PPD at frequencies lower than a chosen frequency threshold and then perform the E-optimal frequency adjustments.

In each iteration the algorithm decides and adjusts one frequency that currently influences the most the lowest eigenvalue of the FIM. This is done numerically using a gradient approach. Therefore, the algorithm iterations consists of the following main parts:
\begin{enumerate}
    \item Estimating the ECM parameters from EIS measurements, computing the FIM and its eigenvalues as described in \ref{subsec:uncertainties} (as discussed, before, the FIM is computed in correspondence of estimated parameters, $\boldsymbol{\theta}_\text{est}$, as $\boldsymbol{\theta}_\text{true}$ is unavailable).
    \item Finding the most suitable candidate frequency that influences the most the lowest eigenvalue of the FIM - this is done by small perturbation of each candidate frequency and finding the highest sensitivity of the perturbation on the lowest eigenvalue of FIM.
    \item Optimal adjustment of the chosen candidate frequency using a gradient approach - chosen frequency is then further perturbed to improve the lowest eigenvalue of FIM until there is no further improvement.
    \item Re-measuring the impedance at adjusted frequency and modifying the set of frequencies.
\end{enumerate}
 The detailed pseudo-code of the algorithm is presented in \cite{sovljanski_use_2024}.

\section{Results and Discussion}\label{sec:numerical_study}

In this section, we show and discuss the results of a numerical study performed on ECM with eleven parameters contained in the parameter vector: 
\begin{equation}
    \small\boldsymbol{\theta} = \left[R_s, Q_{HF}, \phi_{LF}, R_1, Q_1, \phi_1, R_2, Q_2, \phi_2, Q_{LF}, \phi_{LF}\right]^\top.
\end{equation}
The true parameters' values used in this study are obtained by estimating them from real EIS measurements performed on a Li-ion 5 Ah cell SLPB 11543140H5 by Kokam Ltd in two different states: (a) 25~$^\circ$C and $80\%$ SoC and (b) at 15~$^\circ$C and $20\%$ SoC. The parameters are listed in Table~\ref{table:curr_rect_params}. The generated EIS data is corrupted by white noise with maximum relative error of $1\%$ in magnitude and maximum absolute error of $1^\circ$ in phase, corresponding to commercial instruments' standard EIS precision (e.g., \cite{gamry_accuracy_nodate}). 
For the initial EIS scan, frequencies are log-spaced between the starting and final frequencies of interest, $f_{start} = 10$ kHz and $f_{end} = 0.01$ Hz, with $\text{PPD} = 10$.

\begin{table}[h!]
    \centering
    \caption{ECM circuit parameters values.}
    \label{table:curr_rect_params}
    \begin{tabular}{
        c
        c
        S[table-format=1.3e-3]
        S[table-format=1.3e-3]
    }
        \toprule
        $\boldsymbol{\theta}$ & Unit & {State (a)} & {State (b)} \\
        \midrule
        $R_s$       & $\Omega$                            & 1.937e-3 & 2.017e-3 \\
        $Q_{HF}$    & $\Omega\cdot \text{s}^{-\phi_{HF}}$ & 1.132e7 & 1.020e7 \\
        $\phi_{HF}$ & $-$                                 & -9.845e-1 & -9.845e-1 \\
        $R_1$       & $\Omega$                            & 2.409e-3 & 9.535e-3 \\
        $Q_1$       & $\Omega\cdot \text{s}^{-\phi_{1}}$  & 4.715e0 & 8.307e0 \\
        $\phi_1$    & $-$                                 & 6.618e-1 & 5.698e-1 \\
        $R_2$       & $\Omega$                            & 3.273e-3 & 2.647e-2 \\
        $Q_2$       & $\Omega\cdot \text{s}^{-\phi_{2}}$  & 6.419e0 & 6.497e0 \\
        $\phi_2$    & $-$                                 & 9.347e-1 & 9.546e-1 \\
        $Q_{LF}$    & $\Omega\cdot \text{s}^{-\phi_{LF}}$ & 8.585e2 & 6.250e2 \\
        $\phi_{LF}$ & $-$                                 & 5.553e-1 & 5.356e-1 \\
        \bottomrule
    \end{tabular}
\end{table}

\subsection{Case Study}

\begin{figure*}[h!]
    \centering
    \begin{subfigure}[b]{\textwidth}
        \centering
        \begin{tikzpicture}
            \begin{semilogxaxis}[
                axis x line=middle,
                axis y line=none,
                xlabel={$f$ (Hz)},
                xlabel style={at={(axis description cs:1.05,0)}},
                xtick={10^-2,10^-1,10^0,10^1,10^2,10^3,10^4},
                xticklabels={$10^{-2}$,$10^{-1}$,$10^0$,$10^1$,$10^2$,$10^3$,$10^4$},
                scatter/classes={a={mark=o,draw=blue}},
                width=0.8\textwidth, height=3cm,
                enlargelimits=true,
                xmax=1e4 
            ]
        
            \addplot[only marks, mark options={fill=red}, draw=red]
            coordinates {
                (0.01,0) (0.01778,0) (0.03162,0) (0.05623,0) (0.1,0)
            };
        
            \addplot[scatter,only marks,scatter src=explicit symbolic]
            coordinates {
                 (0.1292,0) (0.1668,0) (0.2154,0) (0.2783,0) (0.3594,0) (0.4642,0) (0.5995,0) (0.7743,0) (1,0)
                (1,0) (1.2915,0) (1.6681,0) (2.1544,0) (2.7826,0) (3.5938,0) (4.6416,0) (5.9948,0) (7.7426,0) (10,0)
                (10,0) (12.9155,0) (16.6810,0) (21.5443,0) (27.8256,0) (35.9381,0) (46.4159,0) (59.9484,0) (77.4264,0) (100,0)
                (100,0) (129.1549,0) (166.8101,0) (215.4435,0) (278.2559,0) (359.3814,0) (464.1589,0) (599.4843,0) (774.2637,0) (1000,0)
                (1000,0) (1291.5497,0) (1668.1005,0) (2154.4347,0) (2782.5594,0) (3593.8137,0) (4641.5888,0) (5994.8425,0) (7742.6370,0) (10000,0)
            };
        
            \end{semilogxaxis}
        \end{tikzpicture}
        \caption{}
        \label{subfig:Logspaced_freq_a}
    \end{subfigure}

    \vspace{1pt} 

    \begin{subfigure}[b]{\textwidth}
        \centering
        \begin{tikzpicture}
            \begin{semilogxaxis}[
                axis x line=middle,
                axis y line=none,
                xlabel={$f$ (Hz)},
                xlabel style={at={(axis description cs:1.05,0)}},
                xtick={10^-2,10^-1,10^0,10^1,10^2,10^3,10^4},
                xticklabels={$10^{-2}$,$10^{-1}$,$10^0$,$10^1$,$10^2$,$10^3$,$10^4$},
                scatter/classes={a={mark=o,draw=blue}},
                width=0.8\textwidth, height=3cm,
                enlargelimits=true,
                xmax=1e4 
            ]
        
            \addplot[only marks, mark options={fill=red}, draw=red]
            coordinates {
                (0.01,0) (0.01778,0) (0.03162,0) (0.05623,0) (0.1,0)
            };
        
            \addplot[only marks, mark options={fill=red}, draw=red]
            coordinates {
             (0.17783,0) (0.31623,0) (0.56234,0) (1,0)
            };
        
            \addplot[scatter,only marks,scatter src=explicit symbolic]
            coordinates {
                (1.2915,0) (1.6681,0) (2.1544,0) (2.7826,0) (3.5938,0) (4.6416,0) (5.9948,0) (7.7426,0) (10,0)
                (10,0) (12.9155,0) (16.6810,0) (21.5443,0) (27.8256,0) (35.9381,0) (46.4159,0) (59.9484,0) (77.4264,0) (100,0)
                (100,0) (129.1549,0) (166.8101,0) (215.4435,0) (278.2559,0) (359.3814,0) (464.1589,0) (599.4843,0) (774.2637,0) (1000,0)
                (1000,0) (1291.5497,0) (1668.1005,0) (2154.4347,0) (2782.5594,0) (3593.8137,0) (4641.5888,0) (5994.8425,0) (7742.6370,0) (10000,0)
            };
        
            \end{semilogxaxis}
        \end{tikzpicture}
                \caption{}
        \label{subfig:Logspaced_freq_b}
    \end{subfigure}
    \caption{An example of log-spaced frequency points from $f_{start} = 10^4$~Hz to $f_{end} = 0.01$~Hz with $\text{PPD} = 10$ for every decade, except for decades below the defined frequencies (a) $f = 0.1$~Hz and (b) $f = 1$~Hz, which have $\text{PPD} = 5$.}
    \label{fig:Logspaced_freq}
\end{figure*}
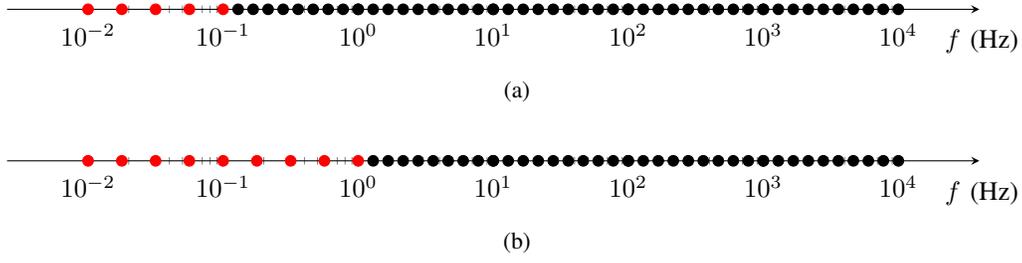

\begin{figure*}[h!]
\centering
\begin{subfigure}{.5\textwidth}
  \centering
  \includegraphics[scale = 0.45]{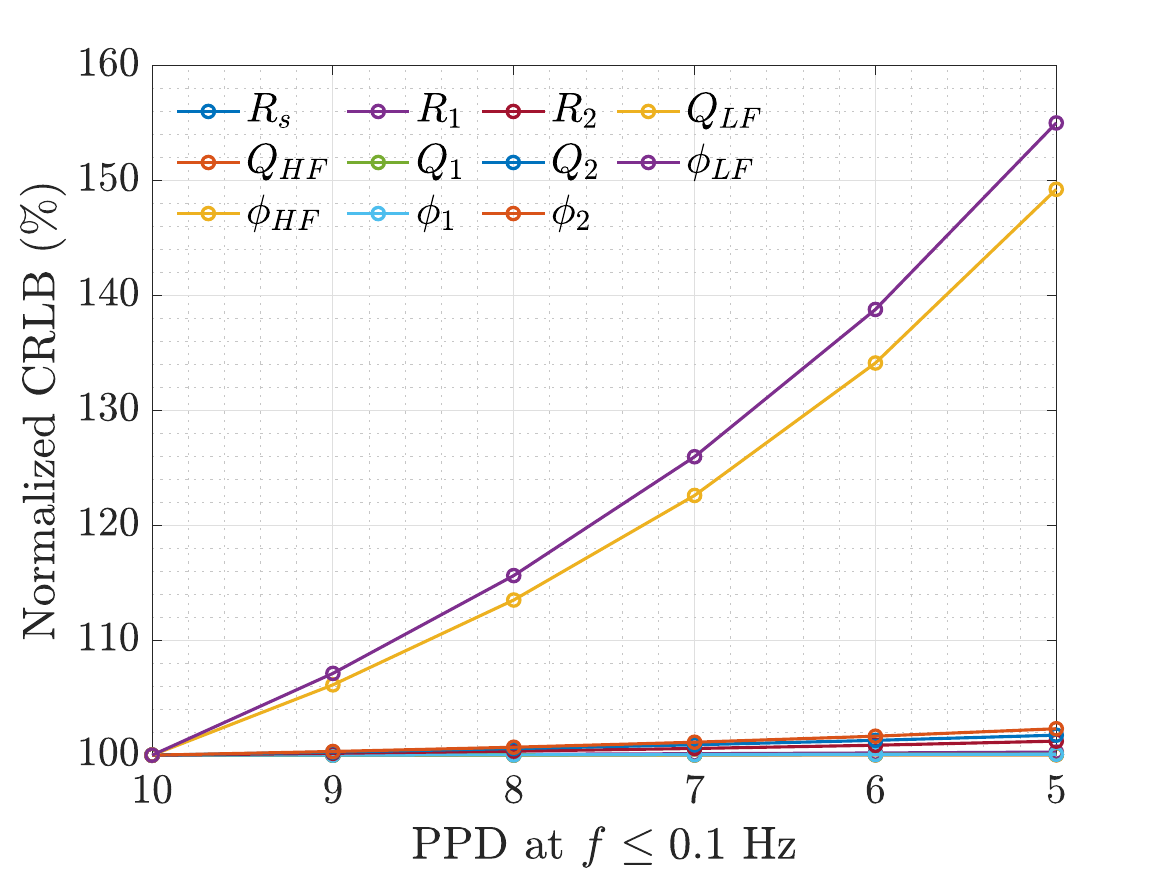}
  \caption{SoC = 80\%, $T = 25 ^\circ$C}
  \label{fig:Influence_of_PPD_reduction_Case1_01Hz}
\end{subfigure}%
\begin{subfigure}{.5\textwidth}
  \centering
  \includegraphics[scale = 0.45]{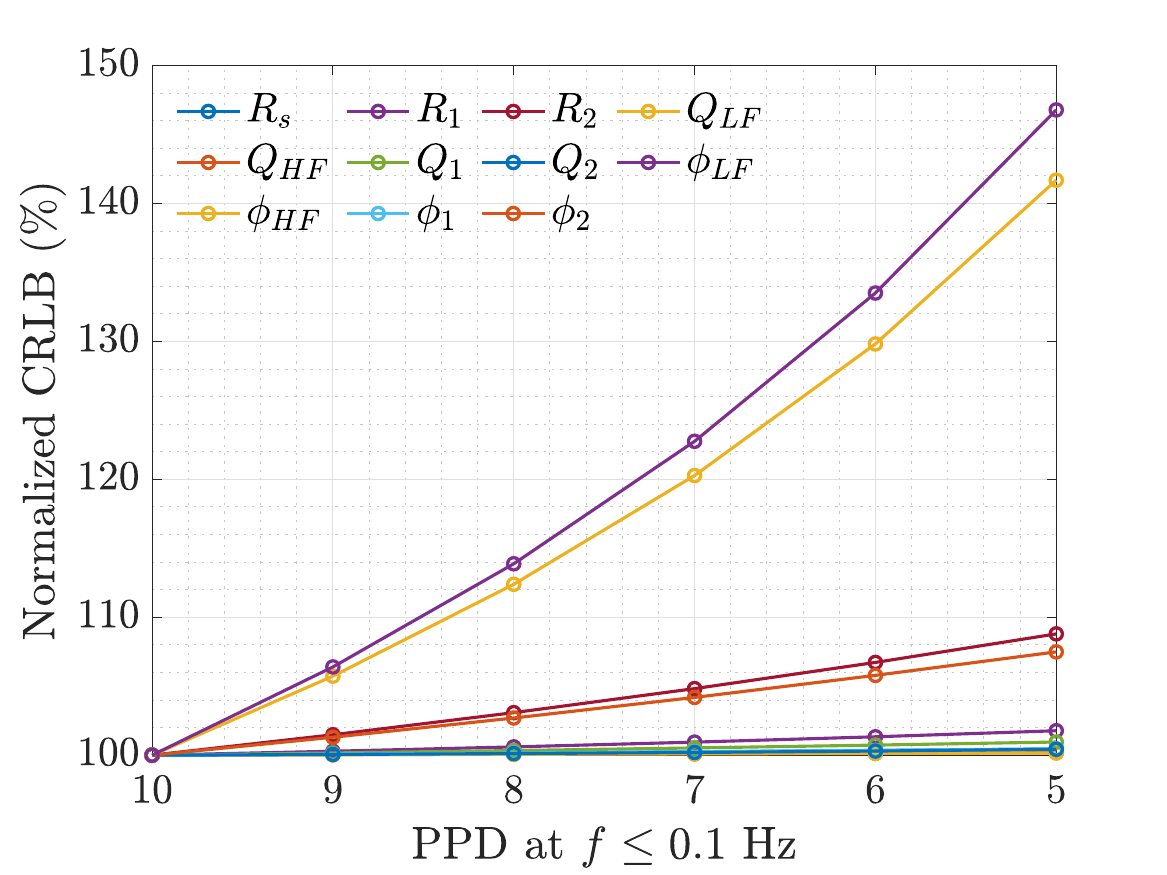}
  \caption{SoC = 20\%, $T = 15 ^\circ$C}
  \label{fig:Influence_of_PPD_reduction_Case2_01Hz}
\end{subfigure}
\caption{Minimum theoretical variance (CRLB) for each of the ECM parameters for different numbers of PPD, i.e., $\text{PPD} \in \{5, 6, \ldots, 10\}$, for frequencies lower than 0.1 Hz.}
\label{fig:Influence_of_PPD_reduction_on_CRLB_01Hz}
\end{figure*}
\begin{figure*}[h!]
\centering
\begin{subfigure}{.5\textwidth}
  \centering
  \includegraphics[scale = 0.45]{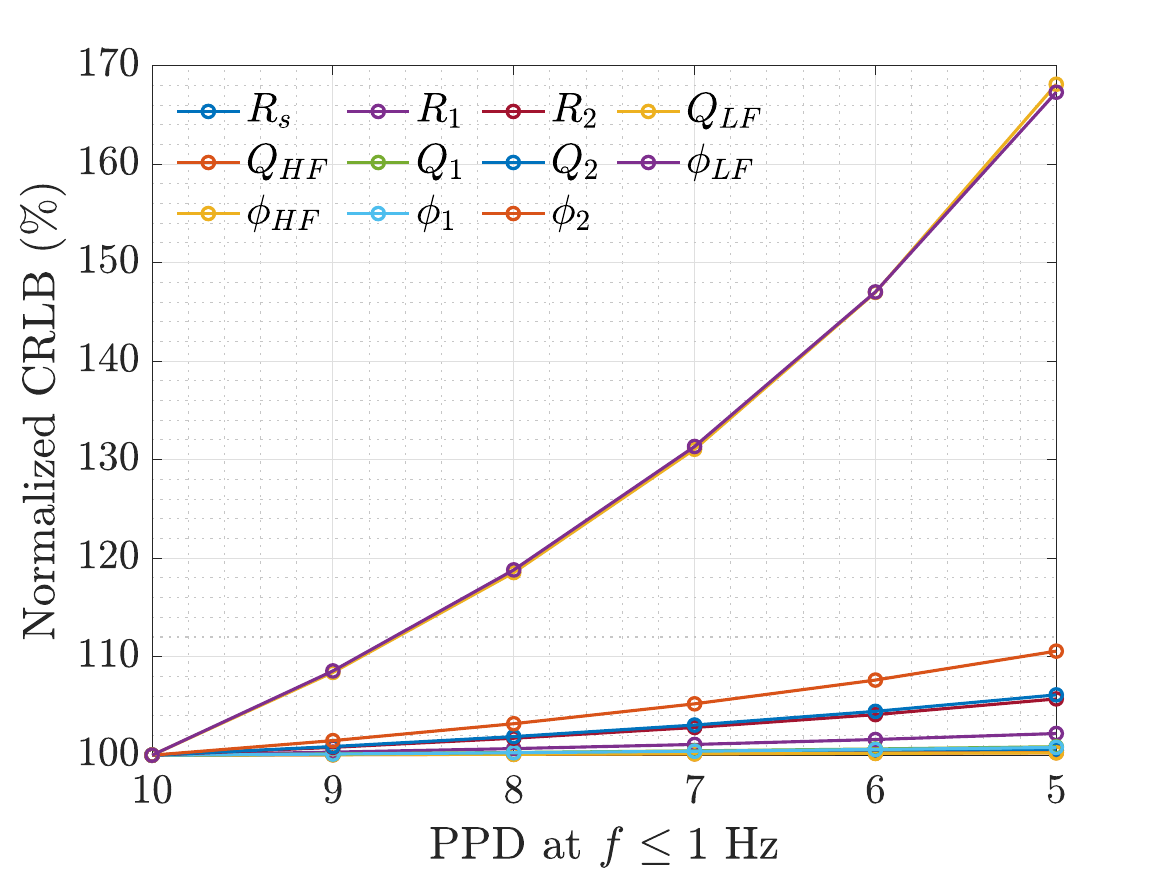}
  \caption{SoC = 80\%, $T = 25 ^\circ$C}
  \label{fig:Influence_of_PPD_reduction_Case1_1Hz}
\end{subfigure}%
\begin{subfigure}{.5\textwidth}
  \centering
  \includegraphics[scale = 0.45]{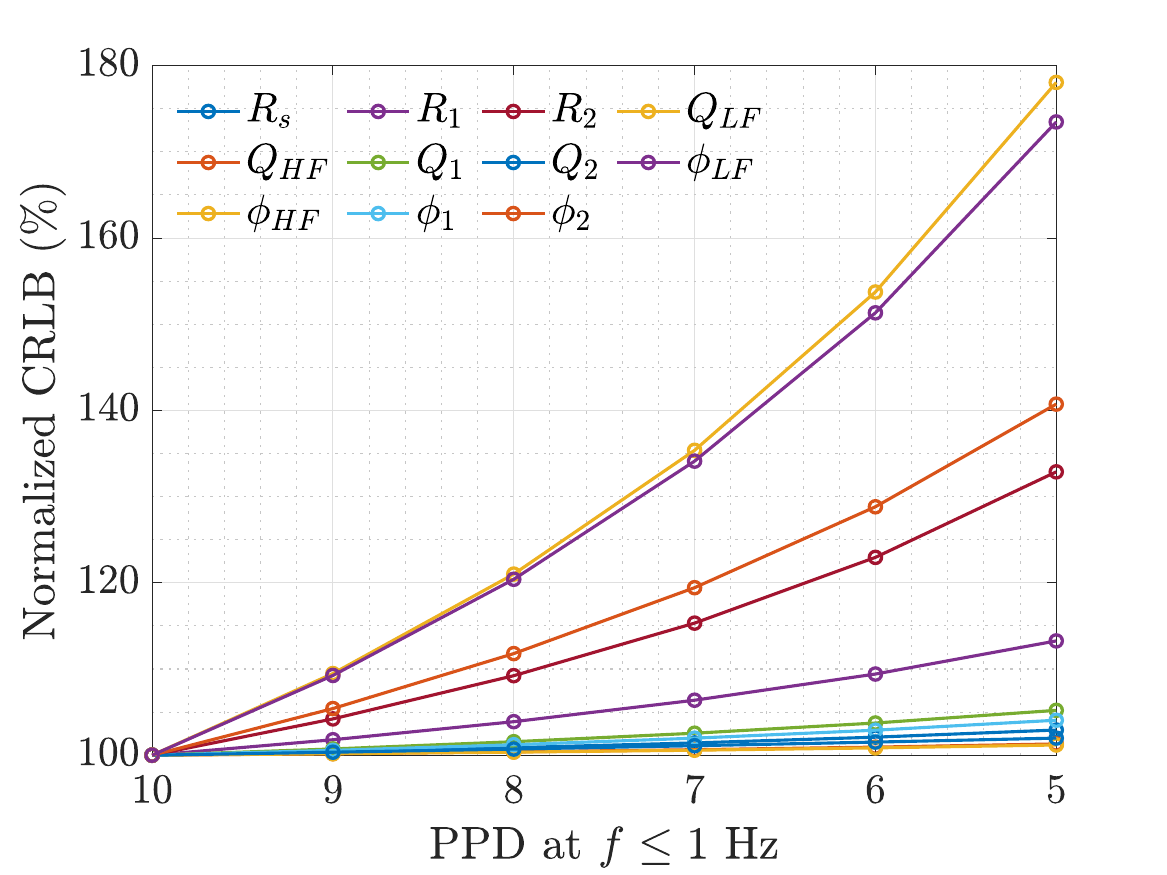}
  \caption{SoC = 20\%, $T = 15 ^\circ$C}
  \label{fig:Influence_of_PPD_reduction_Case2_1Hz}
\end{subfigure}
\caption{Minimum theoretical variance (CRLB) for each of the ECM parameters for different numbers of PPD, i.e., $\text{PPD} \in \{5, 6, \ldots, 10\}$, for frequencies lower than 1 Hz.}
\label{fig:Influence_of_PPD_reduction_on_CRLB_1Hz}
\end{figure*}

\begin{figure*}[h!]
\centering
\begin{subfigure}{.5\textwidth}
  \centering
  \includegraphics[scale = 0.45]{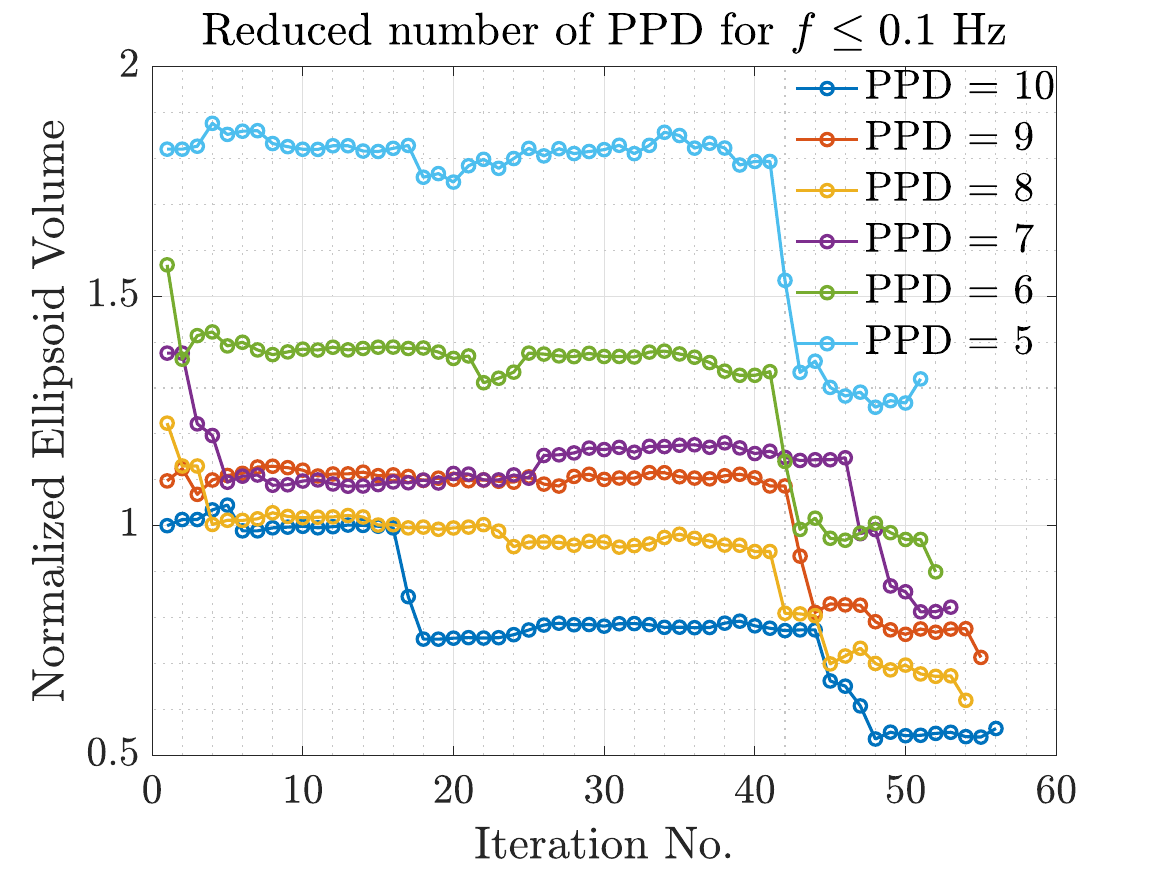}
  \caption{SoC = 80\%, $T = 25 ^\circ$C}
  \label{fig:Ellipsoid_Case1_01Hz}
\end{subfigure}%
\begin{subfigure}{.5\textwidth}
  \centering
  \includegraphics[scale = 0.45]{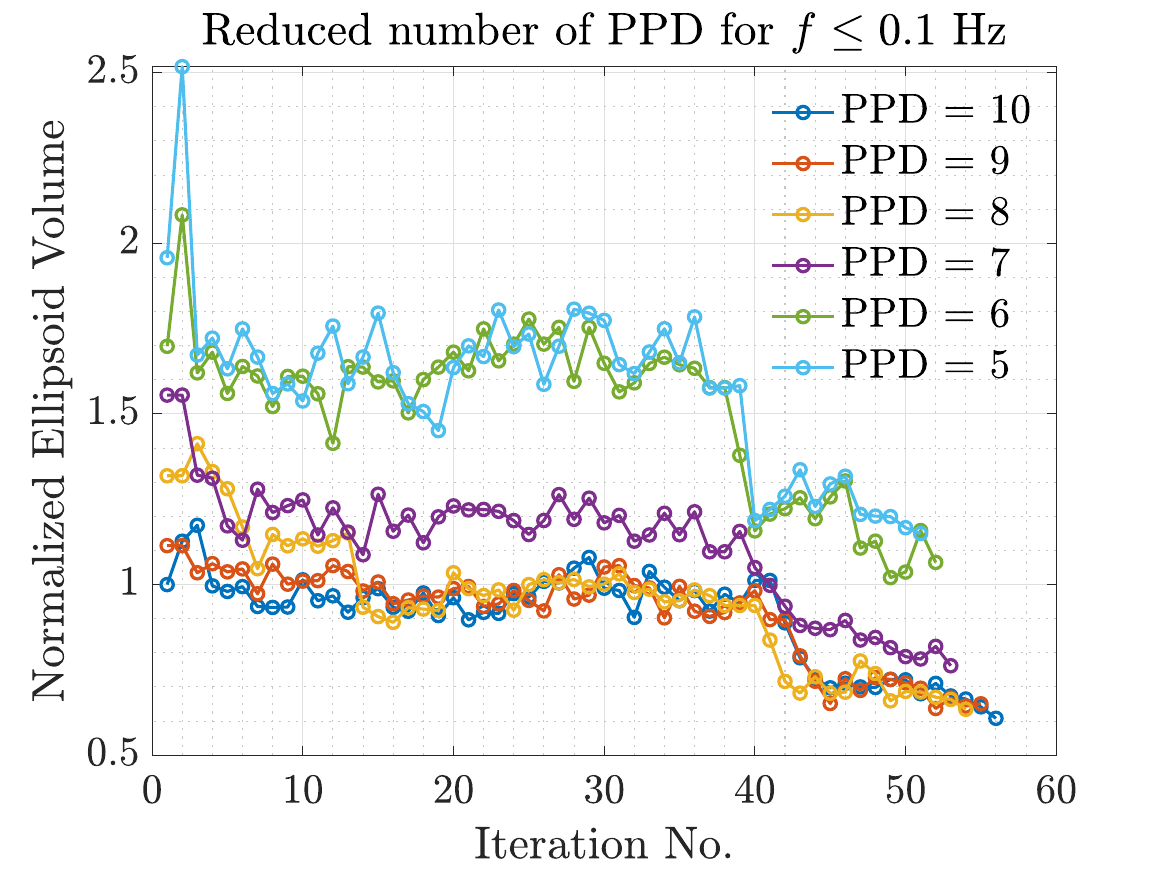}
  \caption{SoC = 20\%, $T = 15 ^\circ$C}
  \label{fig:Ellipsoid_Case2_01Hz}
\end{subfigure}
\caption{Evolution of normalized ellipsoid volume over iterations of the algorithm for optimal frequency adjustments via E-optimal design, starting from a log-spaced frequency set with a reduced number of PPD at frequencies $f \leqslant 0.1$~Hz, for two different cell states.}

\label{fig:Ellipsoids_01Hz}
\end{figure*}

\begin{figure*}[h!]
\centering
\begin{subfigure}{.5\textwidth}
  \centering
  \includegraphics[scale = 0.45]{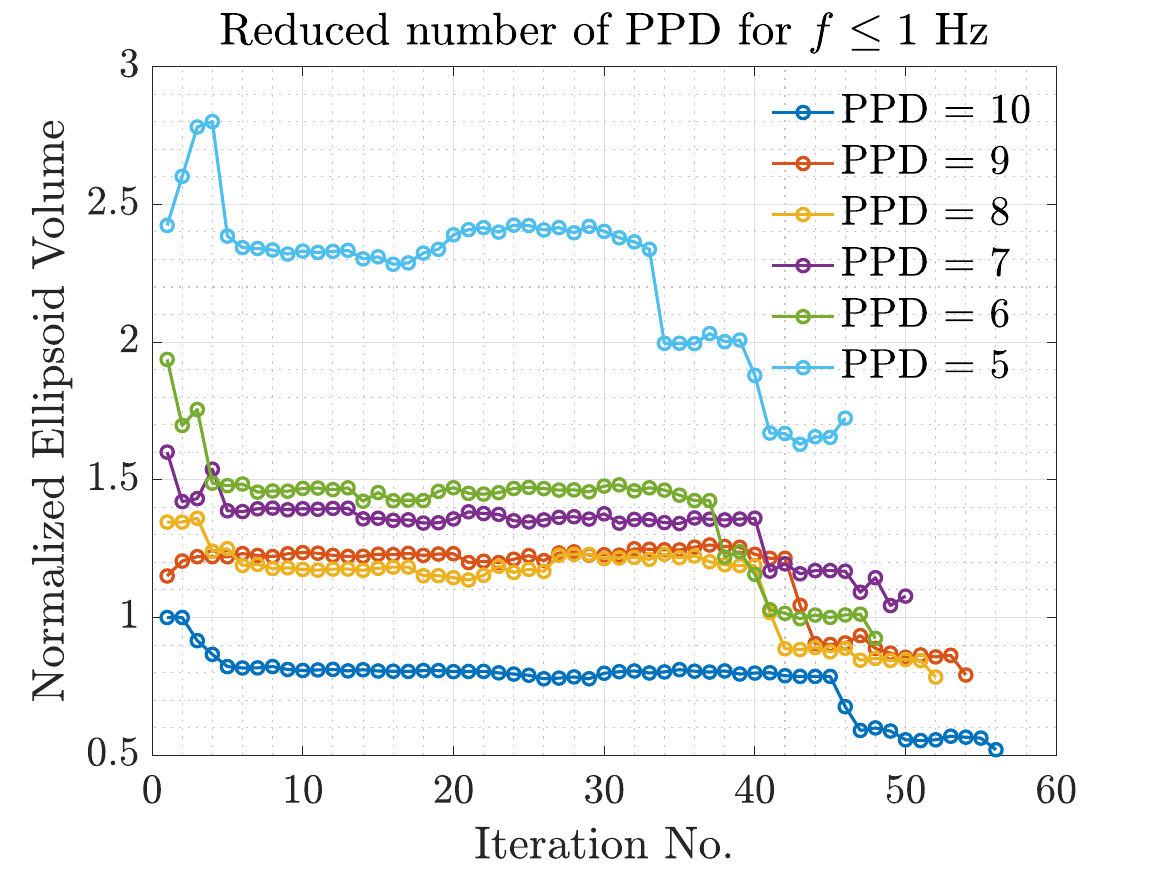}
  \caption{SoC = 80\%, $T = 25 ^\circ$C}
  \label{fig:Ellipsoid_Case1_1Hz}
\end{subfigure}%
\begin{subfigure}{.5\textwidth}
  \centering
  \includegraphics[scale = 0.45]{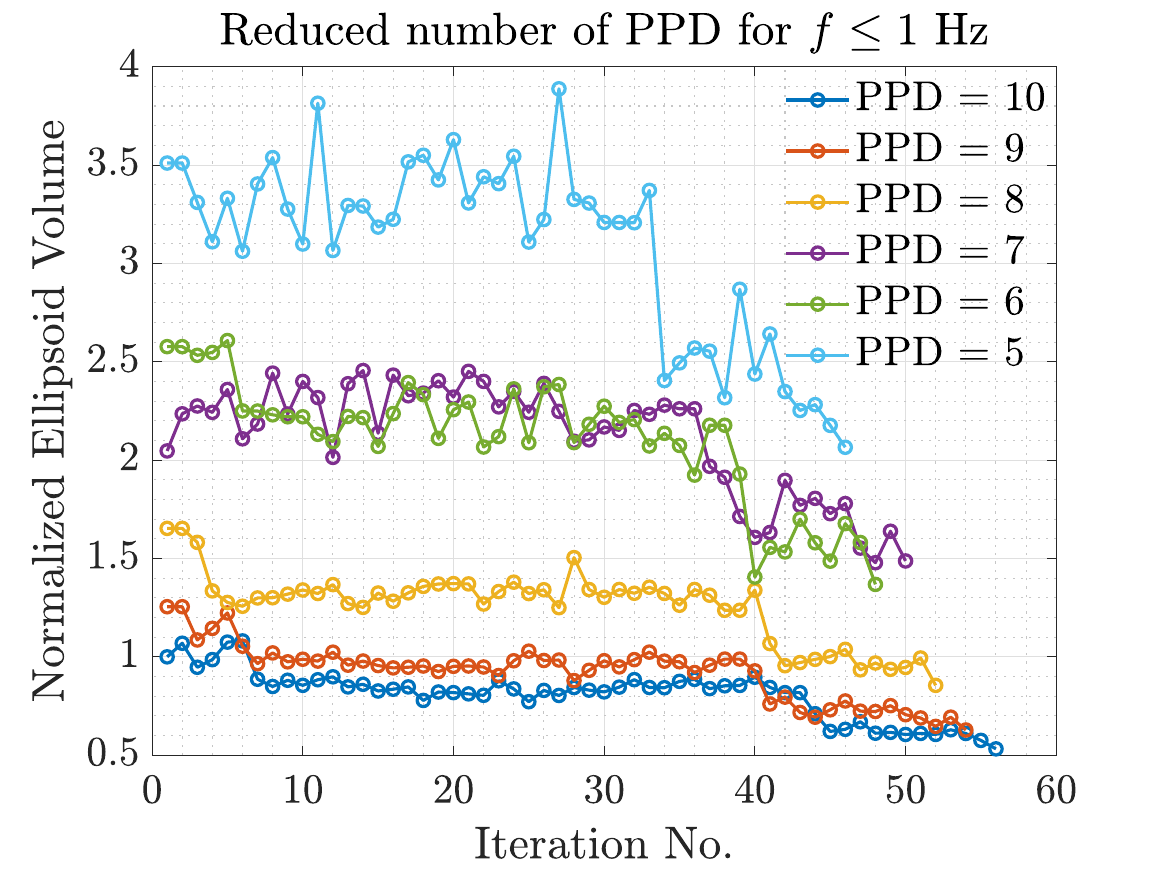}
  \caption{SoC = 20\%, $T = 15 ^\circ$C}
  \label{fig:Ellipsoid_Case2_1Hz}
\end{subfigure}
\caption{Evolution of normalized ellipsoid volume over iterations of the algorithm for optimal frequency adjustments via E-optimal design, starting from a log-spaced frequency set with a reduced number of PPD at frequencies $f \leqslant 1$~Hz, for two different cell states.}
\label{fig:Ellipsoids_1Hz}
\end{figure*}

Removing a certain number of frequencies per decade at lower frequencies significantly reduces the total experimental time but affects the variances of the ECM parameters, particularly those dominant in this frequency region. Here, we study two cases where the number of PPD is reduced for frequencies: ($i$) $f \leqslant 0.1$ Hz and ($ii$) $f \leqslant 1$ Hz. As an example, Fig. \ref{subfig:Logspaced_freq_a} and \ref{subfig:Logspaced_freq_b} illustrate log-spaced frequency sets from $10^{-2}$~Hz to $10^4$~Hz with the default number of $\text{PPD} = 10$, except for the lowest and two lowest frequency decades, respectively, where the number of $\text{PPD}$ is halved.

The impact of the reduction of the number of PPD to $\text{PPD}\in\left\{5,\dots,9\right\}$ for cases ($i$) and ($ii$) is illustrated in Fig.~\ref{fig:Influence_of_PPD_reduction_on_CRLB_01Hz} and \ref{fig:Influence_of_PPD_reduction_on_CRLB_1Hz}, respectively, where for two considered states of the cell we show the computed CRLB for every ECM parameter, normalized with the corresponding parameters' CRLB values for the highest number of $ \text{PPD} = 10$. 
As expected, decreasing the number of PPD at LF leads to worsening (increase) in the least possible variance for all parameters. This increase is most pronounced for the LF CPE parameters. For cell's state (a), reducing the number of PPD from 10 to 5 results in an increase in CRLB of $\theta_{10} = Q_{LF}$ and $\theta_{11} = \phi_{LF}$ by $49.2\%$ and $55\%$ for case ($i$), and by $68.1\%$ and $67.3\%$ for case ($ii$). In contrast, the increase in variance is least evident for HF parameters $\theta_1 = R_s$, $\theta_2 = Q_{HF}$, and $\theta_3 = \phi_{HF}$, which show less than $2\%$ increase in both cases.
Similar holds for the increase in CRLB for $ Q_{LF}$ and $\phi_{LF}$, for the cell's state (b). However, here the increase in CRLB becomes also evident in other parameters, especially the ones of the second Zarc element, $R_2$ and $\phi_2$, with CRLB increase of around $8.8\%$ and $7.5\%$ in case $(i)$ and even $32.9\%$ and $40.7\%$ in case $(ii)$, respectively.

These plots can be used as indicators of the trade-off between the number of PPD at frequencies below a chosen threshold and the uncertainty of the estimated ECM parameters. By simply reducing the number of PPD (without any other frequency adjustments) and thereby reducing the total experimental time, the user accepts increasing the variance (lowering the accuracy) of the estimated ECM parameters by a certain percentage, which can be read from the plot. The plots also show that this reduction is dependent on the cell's state. As discussed, in case ($ii$), reducing the number of PPD has a greater impact on the variance of the estimated MF parameters $R_2$ and $Q_2$ compared to case ($i$).

Fig.~\ref{fig:Ellipsoids_01Hz} and \ref{fig:Ellipsoids_1Hz} show the evolution of the normalized volume of the global uncertainty ellipsoid through the iterations of frequency adjustments using the E-optimal design. The first point of each plot (i.e., at iteration no. 1, for different numbers of PPD) shows how the overall uncertainty increases by reducing the number of PPD at $f \leqslant 0.1$~Hz (Fig. \ref{fig:Ellipsoid_Case1_01Hz} and \ref{fig:Ellipsoid_Case2_01Hz}) and at $f \leqslant 1$~Hz (Fig. \ref{fig:Ellipsoid_Case1_1Hz} and \ref{fig:Ellipsoid_Case2_1Hz}) for both cell states (a) and (b). The volume of each ellipsoid is normalized by the volume of the ellipsoid corresponding to the overall uncertainty of ECM parameters estimated by measuring at the full log-spaced frequency, i.e., without any reduction in the number of PPD and before frequency adjustments.

In the majority of instances, reducing the number of PPD below $0.1$ Hz or $1$ Hz and then adjusting the frequencies leads to a reduced ellipsoid volume to 1 p.u or lower. This is the case whenever the horizontal line indicating the normalized ellipsoid volume equal to 1 p.u at some point crosses the ellipsoid volume evolution plot with a reduced number of PPD. Therefore, thanks to the more optimal frequency distribution at which we perform the EIS measurements, we can achieve the same global uncertainty with fewer measurement points. For cell's state (a), this is indeed confirmed for $\text{PPD} \in \left\{6,7,8,9\right\}$ for case ($i$) and for $\text{PPD} \in \left\{6,8,9\right\}$ for case ($ii$), whereas for cell's state (b), the same global uncertainty can be achieved by reducing the number of PPD to at most $\text{PPD} = 7$ in case ($i$) and to at most $\text{PPD} = 8$ in case ($ii$).

The total experimental time to perform the EIS at the adjusted frequency set with the reduced number of PPD below a certain frequency is desired to be lower compared to the log-spaced frequencies without the reduction of the number of PPD. However, due to frequency adjustments, especially the LF points, it is very likely that the algorithm will decide to decrease the frequency of certain LF points. This is because, for LF ECM parameters, the lower the frequency, the more information is obtained about the parameters due to more negligible interference with other parts of the spectrum. As a consequence, these adjustments towards lower frequency points have an impact on experimental time. To control this, the user can decide to restrict the frequency adjustments, for instance, by adding a constraint on the total experimental time budget or rejecting to decrease the frequencies which are lower than the pre-defined threshold.

\begin{table}[h!]
    \centering
    \caption{Comparison of change of the uncertainty ellipsoid volume and total EIS experimental time for Cases ($i$) and ($ii$) and two considered cell's states.}
    \begin{subtable}{0.45\textwidth}
        \centering
        \label{tab:sub_first}
        \begin{tabular}{
            c
            S[table-format=2.2] 
            S[table-format=2.2] 
            S[table-format=2.2] 
            S[table-format=2.2]
        }
            \toprule
             & \multicolumn{2}{c}{Case ($i$)} & \multicolumn{2}{c}{Case ($ii$)} \\ 
            \cmidrule(lr){2-3} \cmidrule(lr){4-5}
            PPD & {$\Delta V$ (\%)} & {$\Delta t_{tot}$ (\%)} & {$\Delta V$ (\%)} & {$\Delta t_{tot}$ (\%)}  \\ 
            \midrule
$9$   & -28.70        & -9.45        & -20.88        & -13.61        \\
$8$   & -38.04        & -1.48        & -21.64        & -7.18         \\
$7$   & -17.74        & -6.18        & 7.78          & -18.85        \\
$6$   & -10.06        & -27.01       & -7.57         & -15.53        \\
$5$   & 31.98         & -42.63       & 72.40         & -43.24       \\
            \bottomrule
        \end{tabular}
        \vspace{3pt}
        \caption{SoC = 80\%, $T = 25\,^\circ$C}
        \vspace{5pt}
    \end{subtable}
    \begin{subtable}{0.45\textwidth}
        \centering
        \label{tab:sub_second}
        \begin{tabular}{
            c
            S[table-format=2.2] 
            S[table-format=2.2] 
            S[table-format=2.2] 
            S[table-format=2.2]
        }
            \toprule
             & \multicolumn{2}{c}{Case ($i$)} & \multicolumn{2}{c}{Case ($ii$)} \\ 
            \cmidrule(lr){2-3} \cmidrule(lr){4-5}
            PPD & {$\Delta V$ (\%)} & {$\Delta t_{tot}$ (\%)} & {$\Delta V$ (\%)} & {$\Delta t_{tot}$ (\%)}  \\ 
            \midrule
$9$   & -34.94        & 1.03         & -37.19        & 4.67          \\
$8$   & -36.54        & 18.46        & -14.50        & -6.42         \\
$7$   & -23.79        & -9.48        & 48.74         & -29.91        \\
$6$   & 6.50          & -23.87       & 36.82         & -28.78        \\
$5$   & 14.96         & -30.58       & 106.54        & -45.07 \\
            \bottomrule
        \end{tabular}
        \vspace{3pt}
        \caption{SoC = 20\%, $T = 15\,^\circ$C}
    \end{subtable}
    \label{tab:Case_study_V_and_ttot}
\end{table}

\begin{figure*}[h!]
    \centering
    \includegraphics[scale = 0.41]{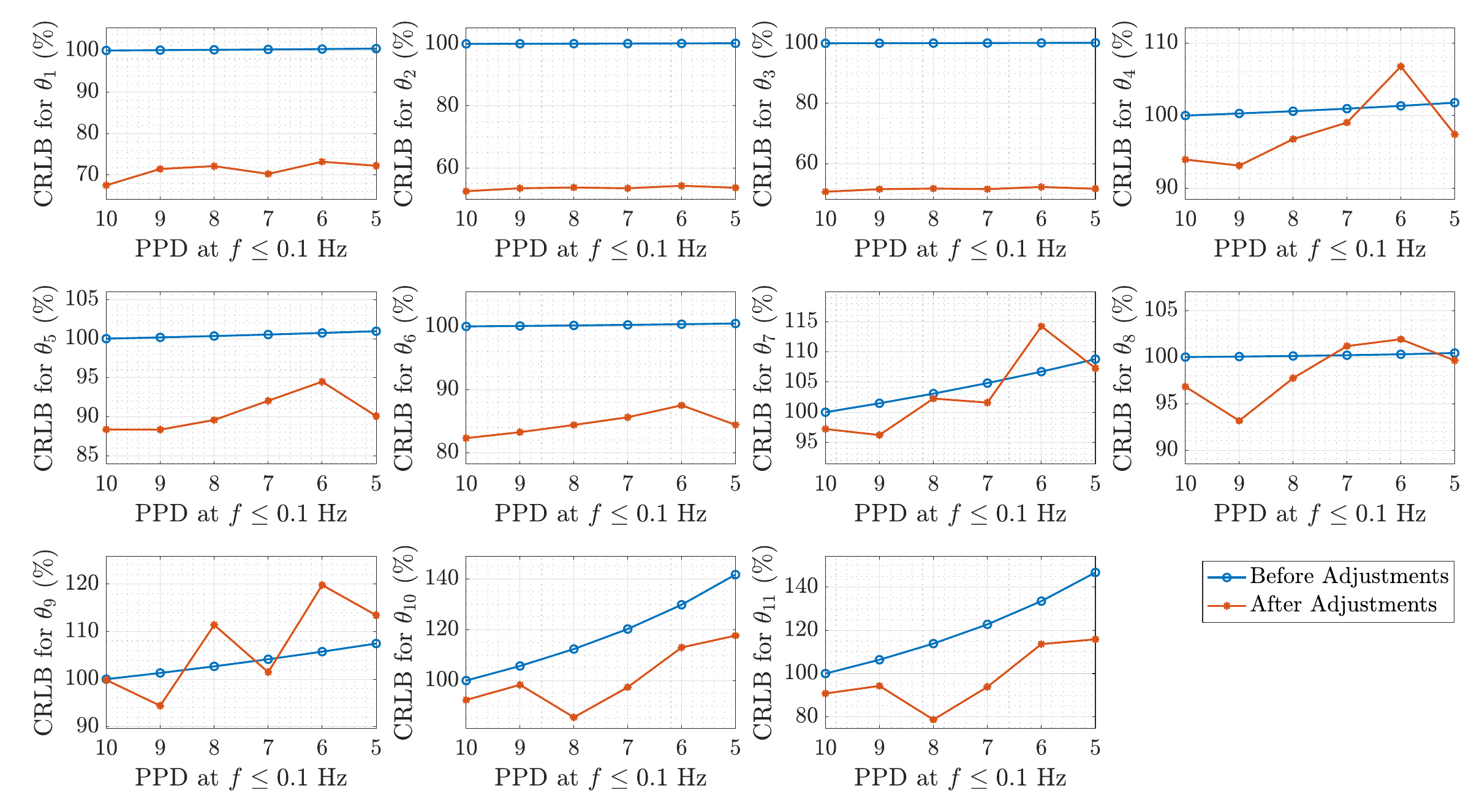}
    \caption{CRLB for each ECM parameter before and after frequency adjustments with the reduced number of $\text{PPD}\in\left\{5,\dots,9\right\}$ at $f\leqslant 0.1$~Hz, for cell's state (b) SoC = 20\%, $T = 15\,^\circ$C. The values are normalized using the CRLB for each parameter corresponding to the log-spaced frequencies with $\text{PPD}=10$.}
    \label{fig:CRLBs_for_each}
\end{figure*}
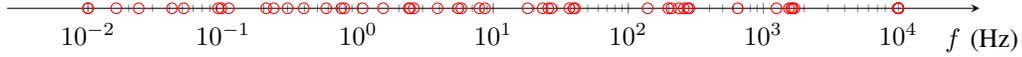
\begin{figure*}[h!]
        \centering
        \begin{tikzpicture}
            \begin{semilogxaxis}[
                axis x line=middle,
                axis y line=none,
                xlabel={$f$ (Hz)},
                xlabel style={at={(axis description cs:1.05,0)}},
                xtick={10^-2,10^-1,10^0,10^1,10^2,10^3,10^4},
                xticklabels={$10^{-2}$,$10^{-1}$,$10^0$,$10^1$,$10^2$,$10^3$,$10^4$},
                width=0.8\textwidth, height=3cm,
                enlargelimits=true,
                xmax=1e4 
            ]
            \addplot[only marks, mark=o, draw=red, mark options={fill=red}]
            coordinates {
                (0.01,0) (0.0100000000000000,0) (0.0161582601752391,0) (0.0237170824512628,0) (0.0417742995025150,0)
                (0.0510969051793471,0) (0.110000000000000,0) (0.0968662248761163,0) (0.0917455295460032,0) (0.236987815903507,0)
                (0.208691955165534,0) (0.395319503018509,0) (0.789070101714172,0) (0.749355312898676,0) (0.580697762010845,0)
                (0.300000000000000,0) (2.58309933002977,0) (2.41874577894009,0) (2.36987815903507,0) (1.53040767121392,0)
                (1.07814409914139,0) (7.89070101714172,0) (8.69252162962465,0) (5.80697762010845,0) (5.50000000000000,0)
                (3.87464899504465,0) (39.2003626242014,0) (36.6253897305420,0) (40.3471113320033,0) (26.9536024785347,0)
                (25.5287385848703,0) (17.9845275095682,0) (23.2279104804338,0) (275,0) (258.309933002977,0)
                (283.577091324010,0) (236.987815903507,0) (208.691955165534,0) (197.659751509254,0) (139.247665008383,0)
                (1648.58168837709,0) (1548.52736536225,0) (1700,0) (1614.43708126860,0) (1251.07540290004,0)
                (646.330407009565,0) (10000,0) (10000,0) (10000,0) (10000,0) (10000,0) (10000,0)
            };
            \end{semilogxaxis}
        \end{tikzpicture}
        \caption{Resulting frequency points after adjusting the initial log-spaced frequency set with $\text{PPD} = 7$ at $f\leqslant 0.1$~Hz for cell's state (b) SoC = 20\%, $T = 15\,^\circ$C.}
        \label{fig:final_adjustment}
\end{figure*}
Table~\ref{tab:Case_study_V_and_ttot} shows the percentage change in uncertainty ellipsoid volume, $\Delta V$, and total experimental time, $\Delta t_{tot}$ to perform the EIS at adjusted sets of frequencies starting from initial log-spaced frequencies with a reduced number of PPD for both cases ($i$) and ($ii$) and both considered cell's states (a) and (b). All quantities are normalized by the corresponding ellipsoid volume and total experimental time required for the EIS at the full log-span frequency set with $\text{PPD} = 10$ without reduction of the number of PPD and before any frequency adjustments. Negative change indicates the improvement (reduction) in ellipsoid volume and total experimental time. For case ($i$), the reduction from $\text{PPD} = 10$ to $\text{PPD} = 7$ results in an improvement of both global uncertainty and total experimental time for both cell's states, whereas for case ($ii$), this is achieved when the number of $\text{PPD}$ is reduced to at most $\text{PPD} = 8$.

Fig.~\ref{fig:CRLBs_for_each} shows the normalized CRLB for each ECM parameter before and after adjustments of frequency set with the number of $\text{PPD}\in\left\{5,\ldots,10\right\}$ at $f\leqslant 0.1$~Hz for cell's state (b). In the case of $\text{PPD} = 7$, after executing the algorithm for frequency adjustments, the CRLB for almost all the ECM parameters is improved compared to the log-spaced frequency set with $\text{PPD} = 10$. The reduction of CRLB is even present in LF parameters: $2.6\%$ for $Q_{LF}$ and $6.2\%$ for $\phi_{LF}$. HF parameters $R_s$, $Q_{HF}$ and $\phi_{HF}$ significantly improve their variance by $29.7\%$, $46.6\%$ and $48.5\%$, respectively. Furthermore, parameters $Q_1$ and $\phi_1$ of the first Zarc improve by $8\%$ and $14.4\%$, respectively, whereas the CRLBs for parameters $R_1$, $R_2$, $Q_2$ and $\phi_2$ remain within $2\%$ compared to the full log-spaced scan. The final frequency placement at which these improvements are attained is shown in Fig.~\ref{fig:final_adjustment}.

\section{Conclusion}\label{sec:conclusion}
In this paper, we explored the potential for reducing the frequency set used in EIS when characterizing Li-ion cells. Our case study demonstrates how reducing the number of PPD at lower frequencies affects the minimum theoretical variance of estimated ECM parameters as defined by the CRLB. The LF parameters of the ECM are particularly impacted, with a significant variance increase when the number of PPD at LF is halved. However, by optimally adjusting the reduced frequency set, we show that the global uncertainty, characterized by the uncertainty ellipsoid, can be maintained compared to the full log-scanned EIS measurements while reducing the total experimental time.
Specifically, reducing the number of PPD for frequencies $f \leqslant 0.1$~Hz from $\text{PPD} = 10$ to $\text{PPD} = 7$ and then adjusting the frequency set using the E-optimal design results in improved global uncertainty, characterized by the uncertainty ellipsoid volume, by $17.7\%$ at SoC = 80\% and $T = 25\,^\circ$C, and by $23.8\%$ at SoC = 20\% and $T = 15\,^\circ$C. In these cases, the total experimental time is reduced by $6.2\% $ and $9.5\%$, respectively.
This reduction in the number of PPD at lower frequencies saves experimental time and, together with frequency adjustments, makes it highly applicable when both experimental time and estimated ECM parameters accuracy are critical factors in the experiment design for Li-ion cell characterization.

\section*{Acknowledgment}
This research is carried out in the frame of EPFL – PSA Stationary Storage Study, financed by Stellantis and of the Swiss Circular Economy Model for Automotive Lithium Batteries (CircuBAT) Flagship project, with the financial support of the Swiss Innovation Agency (Innosuisse - Flagship Initiative) (FLAGSHIP PFFS-21-20).

\bibliographystyle{IEEEtran}
\bibliography{bibliography}
\end{document}